\documentclass[aps,superscriptaddress,twocolumn,preprintnumbers,floatfix,showpacs,prl]{revtex4-1}

\usepackage{amsmath,amssymb}
\usepackage{natbib}
\usepackage{booktabs}
\usepackage{multirow}
\usepackage{hyperref}
\usepackage{dcolumn}
\usepackage{braket}
\usepackage{dsfont}
\usepackage{feynmp}
\usepackage{hhline}
\usepackage{graphicx}
\usepackage{mathtools}
\usepackage{xfrac}
\usepackage{paralist}
\usepackage{enumitem}
\usepackage[normalem]{ulem}
\usepackage{color}
\definecolor{darkblue}{rgb}{0.,0.,0.4}
\definecolor{darkred}{rgb}{0.5,0.,0.}

\newcommand{\rb}{\boldsymbol{r}}

\newcommand{\sbb}{\boldsymbol{s}}

\newcommand{\ob}{\boldsymbol{o}}
\newcommand{\eb}{\boldsymbol{e}}

\newcommand{\Eb}{\boldsymbol{E}}
\newcommand{\Mb}{\boldsymbol{M}}

\newcommand{\psib}{\boldsymbol{\psi}}

\newcommand{\beq}{\begin{eqnarray}}
\newcommand{\eeq}{\end{eqnarray}}

\DeclareMathOperator{\sgn}{\mathrm{sgn}}
\begin{document}

\title{Topological Order and Symmetry Anomaly on the Surface of  \\ Topological Crystalline Insulators}
\author{Sungjoon Hong}
\author{Liang Fu}
\affiliation{Department of Physics, Massachusetts Institute of Technology, Cambridge, MA 02139, USA}
\begin{abstract}
We present microscopic Hamiltonians that gap the Dirac fermions on the surface of  topological crystalline insulators (TCIs) protected by reflection symmetry and  create symmetry-preserving states with Abelian topological orders. For TCIs with $n=2$ copies of surface Dirac fermions, we find a fermionic $\mathbb{Z}_4$ topological order with charge-spin self duality and with a novel symmetry action on the anyons, whereby reflections with respect to two parallel planes act differently. We also analyze TCIs with $n=1$ and $4$ copies of Dirac fermions, which taken together covers every nontrivial case in the $\mathbb{Z}_8 \oplus \mathbb{Z}_2$ classification of interacting TCIs. Our work reveals a fundamental difference between reflection and charge-conjugation-time-reversal symmetry anomalies.

\end{abstract}
\maketitle

\noindent{\it Introduction:}  Topological crystalline insulators (TCIs)
 are topological states of matter protected by charge conservation and crystal symmetry such as rotation and reflection \cite{ando2015topological}.
The first material realization of TCI was found in SnTe and related IV-VI semiconductors \cite{hsieh2012topological,tanaka2012experimental,
dziawa2012topological,xu2012observation,okada2013observation}, whose topological nature relies on the reflection symmetry of the rocksalt crystal. Recently, a new type of TCI, protected jointly by glide reflection and time reversal symmetry, was found in KHgSb\cite{wang2016hourglass,alexandradinata2016topological,ma2016experimental}. The richness of crystallography provides a vast opportunity for theoretical and experimental discovery of new classes of TCIs
\cite{shiozaki2014topology,
fang2015new,shiozaki2015z,liu2014topological,parameswaran2013topological,watanabe2015filling,
shiozaki2016topology,chiu2016classification,alexandradinata2015spin,shiozaki2017topological,chang2017mobius,ramamurthy2017electromagnetic}. Systematic studies of TCIs are expected to provide an organizing principle for electronic structures of solids.

While early studies of TCIs are based on topological band theory, whether TCIs are stable against electron interactions is an important theoretical question. Recent works starting with Isobe and Fu \cite{isobe2015theory} have shown that the inclusion of electron interaction alters the classification of TCIs
\cite{song2017topological, morimoto2015breakdown,song2017interaction,watanabe2017topological}. Non-interacting TCIs protected by reflection symmetry $M$ are classified by an integer topological index known as the mirror Chern number ($n_M$) \cite{teo2008surface}, which determines the number of Dirac cones on surfaces that preserve the reflection symmetry. When the interaction is introduced, this integer classification reduces to a $\mathbb{Z}_8$ subgroup \cite{isobe2015theory}, implying that the case of $n_M=0 \mod 8$ is adiabatically connected to a trivial insulator. Moreover, electron interaction enables a new TCI phase that cannot be realized in non-interacting systems \cite{song2017topological}. From these studies,  the complete $\mathbb{Z}_8 \oplus \mathbb{Z}_2$ classification of interacting TCIs protected by charge conservation and reflection symmetry is obtained.

In this work, we introduce microscopic models to systematically study topologically ordered phases on the surface of TCI driven by strong electron interactions.  Recent theoretical breakthroughs  \cite{wang2013gapped,wang2014interacting,wang2014classification,
metlitski2014interaction,chen2014symmetry,
metlitski2015symmetry,senthil2015symmetry} have found that 3D symmetry-protected topological phases (SPTs) may have gapped and symmetry-preserving 2D surface states, which exhibit topological order and support fractional excitations (anyons). Importantly, the relevant symmetry acts on anyons in an anomalous way that cannot be realized in any two-dimensional system \cite{lu2012theory,chen2015anomalous,hermele2016flux,wang2016anomaly}. This theme of surface topological order with symmetry anomaly has been studied in many SPTs with internal symmetries, but to a much lesser extent in TCIs \cite{qi2015anomalous,mross2016anomalous,lake2016anomalies}. Moreover, these studies are mostly based on field theoretic techniques without providing microscopic models. Our work builds on the key idea that surface Dirac fermions in TCIs can be deformed into an array of one-dimensional edge states \cite{isobe2015theory,fulga2016coupled}, and uses the powerful coupled-wire method \cite{teo2014luttinger} to construct  microscopic models that realize topologically ordered surface states and analyze the nontrivial symmetry action on anyons.

 Our most interesting result is for TCIs with $n=2$ copies of surface Dirac fermions, as realized in SnTe. Here we find a gapped, symmetry-preserving surface state with {\it fermionic}  $\mathbb{Z}_4$ topological order. We show that reflection symmetry permutes anyon types in a highly unusual way, and reflections with respect to two parallel mirror planes act differently. 
This anyon permutation property has important consequences for possible phase transitions driven by anyon condensation. We also briefly present the cases of $n=4,1$. We compare our results on reflection symmetry anomaly with previous field-theoretic works on charge-conjugation-time-reversal symmetry anomaly \cite{fidkowski2013non,metlitski2014interaction,potter2016realizing} and discuss their similarities and differences.

\noindent{\it Model:}  Consider a 3D TCI protected by the reflection symmetry $y \rightarrow -y$ which hosts, in the noninteracting limit,  $n=2$ copies of 2D Dirac fermions on the surface parallel to $xy$ plane. The Dirac Hamiltonian for surface states is given by
\beq
H=\sum_{a=1}^2 v_F \int d^2 \rb~ \psi^\dagger_a(\rb)(-i\partial_x \sigma_y +i\partial_y\sigma_x)\psi_a(\rb)
\eeq
where $\psi^\dagger _a=(\psi^\dagger_{a,\uparrow},\psi^\dagger_{a,\downarrow})$, and $\sigma_{x,y,z}$ are $2\times 2$ Pauli matrices associated with electron's spin. Reflection acts jointly on electron's spatial coordinate and spin as follows,
\beq
M_y: \psi^\dagger_{a }(x,y) \rightarrow \psi^\dagger_{a}(x, -y) \sigma_y.
\eeq

Since it is difficult to analyze the non-perturbative effect of strong electron interactions directly, we first deform these 2D Dirac fermions into an array of 1D edge states by introducing a periodically alternating Dirac mass term \cite{isobe2015theory},
\beq\label{mass}
H_m=  \int d\rb \; m(y) [\psi^\dagger_1(\rb) \sigma_z \psi_1(\rb)- \psi^\dagger_2 (\rb) \sigma_z \psi_2( \rb)]
\eeq
where $m(y)=m_0$ for $2j<y<2j+1$ and $-m_0$ for $2j-1<y<2j$.  Since Dirac mass is odd under reflection, our setup is symmetric under reflection. When $m_0$ is sufficiently large, we obtain an array of gapless one-dimensional states localized at Dirac mass domain walls. Each domain wall has left and right movers with opposite mirror eigenvalues due to spin-momentum locking, i.e.,
\beq
M_j:\psi^{R,L}_{j+n}\rightarrow \pm (-1)^{j+n}\psi^{R, L}_{j-n}.
\eeq
where $M_j$ denotes reflection with respect to the wire at $j$. It is important to note that our setup is invariant under $y \rightarrow y+2$, and has two sets of inequivalent reflection planes located at $y=2j$ and $y=2j+1$ respectively.

These one-dimensional states can be conveniently described by Luttinger liquid using density and phase variables
\beq\label{free Hamiltonian}
H_0= \sum_{j}  \frac{v_f}{2\pi}\int dx  (\partial_x \phi_j)^2+ (\partial_x \theta_j)^2
\eeq
where physical electron operators are given by standard Bosonization rule $\psi^{R,L}(x)\sim \kappa \exp(i(\phi(x) \pm \theta(x))$\cite{giamarchi2004quantum}. Commutation relation is given by $[ \partial_x \theta(x), \phi(x')] = i \pi \delta(x-x')$.  In terms of Bosonic variables, symmetry actions are given by
\beq\label{symmetry}
&&M_j: \Big\{
\begin{array}{ll}
\phi_{j+n} \rightarrow \phi_{j-n}+ \pi/2 \\
\theta_{j+n} \rightarrow \theta_{j-n} +(-1)^{j+n+1}\pi/2
\end{array} \\
&&U(1)_c: \phi_j \rightarrow \phi_j +\alpha \label{u1}\label{reflection}
\eeq
where $U(1)_c$ is associated with charge conservation.

\begin{figure}
\includegraphics[scale=0.65]{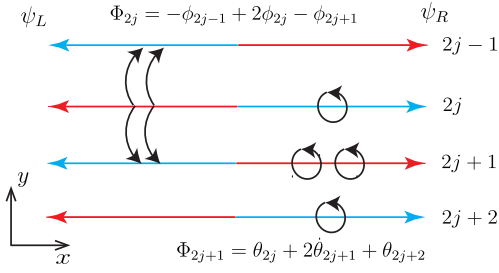}
\caption{\label{fig1}Array of mass domain wall on the surface of a TCI preserves reflection symmetry $y\rightarrow -y$ and creates an array of one-dimensional left- and right-moving states. Red and blue colors label mirror eigenvalues $\pm 1$.
Two types of multi-body electron interactions, schematically shown here and defined in  Eq.(\ref{Hv}), open up an energy gap and leads to a topologically ordered surface state.}
\end{figure}

Following the general approach of coupled-wire construction \cite{mukhopadhyay2001sliding,kane2002fractional,teo2014luttinger,neupert2014wire,meng2015coupled,iadecola2016wire,sahoo2016coupled,patel2016two} , we  introduce two types of three-wire interaction terms to gap the surface states while preserving the $M\times U(1)_c$ symmetry. First we define a new set of bosonic field variables and their conjugates:
\beq
&&\Phi_{2j+1} = \theta_{2j}+2 \theta_{2j+1}+ \theta_{2j+2},~\Theta_{2j+1}=\phi_{2j+1}  \nonumber \\
&&\Phi_{2j} = -\phi_{2j-1}+ 2 \phi_{2j}- \phi_{2j+1},~\Theta_{2j} =  \theta_{2j}
\eeq
with commutation relation given by $[ \partial_x \Theta_j(x), \Phi_{j'}(x')] = 2i  \pi\delta_{jj'} \delta(x-x')$. The interactions we introduce are
\beq\label{Hv}
H_V =\sum_j \int dx  V_1 \cos 2\Phi_{2j+1} + V_2 \cos 2\Phi_{2j}
\eeq
The $V_1$ term describes correlated electron backscattering in three wires ($2j, 2j+1, 2j+2$), while the $V_2$ term describes correlated pair hopping between the two wires ($2j-1, 2j+1$) and the middle one $2j$. 

Note that in addition to the physical charge conservation $U(1)_c$, our Hamiltonian $H=H_0 + H_V$ has an axial $U(1)_a$ symmetry associated with the opposite $U(1)$ transformation of left and right movers
$
\psi^{R,L}_j \rightarrow \exp (\pm i(-1)^j\alpha ) \psi^{R,L}_j
$
or equivalently
\beq\label{axial}
U(1)_a: \theta_j \rightarrow \theta_j+(-1)^j \alpha
\eeq
Since left and right movers carry opposite spins in the $y$ direction, this axial $U(1)$ symmetry is due to the spin $\sigma_y$ conservation in our model.
With both charge and spin $U(1)$ symmetry present, our Hamiltonian exhibits an interesting charge-spin self duality under the transformation given by
\beq
\phi_j \leftrightarrow (-1)^j \theta_{j+1}, ~~ V_1\leftrightarrow V_2
\eeq
which exchanges the spin ($\theta$) and charge ($\phi$) degrees of freedom.

All terms in $H_V$ mutually commute.  
When $V_1$ and $V_2$ are sufficiently large, $H_V$ drives the array of wires to the strong coupling limit, where all $\Phi$ fields are pinned to classical values and an energy gap is created. Although $\langle e^{2i\Phi}\rangle\neq0$, spontaneous symmetry breaking is ruled out as $e^{i\Phi}$s are not local observables. We thus obtain a gapped and symmetry-preserving surface states in TCIs.

\noindent{\it Topological Order:} We now analyze fractional excitations and their statistics in the strong coupling limit. A bulk quasiparticle corresponds to a finite energy soliton configuration of $\Phi$, which connects different minima of cosine potential in $H_V$. Smallest such kinks are $\pi$ kink, which can be conveniently expressed as $\Phi(x)=\Phi_{minimum}+\pi \eta(x-x')$, where $\eta(x)\approx(1+\sgn(x))/2$ is a step function. We will denote the $\pi$-kinks of $\Phi_{4j}$ and $\Phi_{4j-1}$ as $\eb$ and $\ob$, whose energy cost is determined by $V_1$ and $V_2$ respectively. These $\eb$ and $\ob$ excitations are {\it fractional} excitations, because they cannot be created alone by any local physical operators. Non-local operators that create a single $\eb, \ob$ anyons are
\beq\label{creation operator}
e^\dagger\equiv e^{i\theta_{4j}(x)/2},~~o^\dagger\equiv e^{i\phi_{4j-1}(x)/2}
\eeq

Applying a local electron backscattering operator on wire $4j$,  $\psi^\dagger_{4j,L}(x')\psi_{4j,R}(x') = e^{2i\theta_{4j}(x')}$,  results in the transformation $\partial_x \Phi_{4j}(x) \rightarrow \partial_x \Phi_{4j}(x) + 4\pi \delta(x-x')$, and leave all other $\Phi$ fields invariant. This implies that a $4\pi$ kink of $\Phi_{4j}$---a composite of four $\eb$ excitations---is a local quasiparticle. The same applies to $\ob$.
This fact implies the fusion rule
\beq
\eb^4 \sim \ob^4 \sim 1. \label{fusion}
\eeq

Consider another multi-body backscattering operator $(\psi^L_{4j+1})^\dagger (\psi^R_{4j+2}\psi^L_{4j+2})(\psi^R_{4j+3})^\dagger$, which results in the transformation $\Phi_{4j}(x) \rightarrow \Phi_{4j}(x)-\pi\eta(x-x') ,~ \Phi_{4j+4}(x) \rightarrow \Phi_{4j+4}(x)+\pi\eta(x-x')$. The effect of this operator is to move a $\eb$ excitation from position $4j$ to $4j+4$. Similar result holds for $\ob$.  On the other hand, there is {\it no} local bosonic operator that moves a $\pi$ kink from position $j$ to $j\pm 2$.
Therefore, despite the invariance of the microscopic Hamiltonian under translation $j\rightarrow j+2$, $\pi$ kinks at position $4j+2$ and $4j+1$, which we denote  by $\eb'$ and $\ob'$ respectively, are two additional types of fractional excitations distinct from either $\eb$ or $\ob$. They are created by nonlocal operators $e'^\dagger, o'^\dagger$, which can be defined similar to (\ref{creation operator}). The general property that global symmetry can change anyon types also appears in other models \cite{kitaev2006anyons,teo2014unconventional,vijay2015majorana}.

Considering the role of physical electrons helps to understand the full structure of topological excitations. It is straightforward to show that applying the electron operator $\psi_{4j} \sim e^{i(\phi \pm \theta)}$ to the ground state creates $\ob$, $\ob'$, $\eb^2$ excitations. This implies the following fusion rules among $\eb,\eb',\ob,\ob'$ anyons:
\beq\label{enumeration}
\begin{dcases}
\ob' \sim \eb^2 \times \ob^3 \times \psib \\
\eb' \sim \eb^3 \times \ob^2 \times \psib
\end{dcases}
\eeq
We thus conclude that the full set of anyons in our system can be divided into two classes. The first class is generated from the two ``root'' anyons $\eb$ and $\ob$, while the second class is obtained by attaching physical electrons $\psib$ to the first class. The fact that physical electrons should be included in the counting of topological excitations, known as fermion parity grading, is a universal property of topologically ordered states in fermion systems \cite{vijay2015majorana}.

\begin{figure}
\includegraphics[scale=0.75]{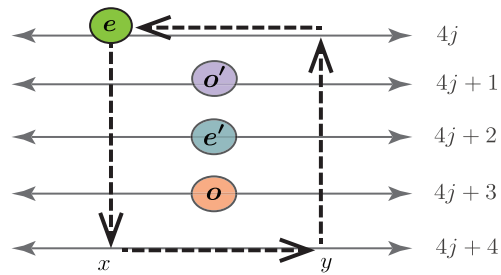}
\caption{\label{fig2}Braiding an $\eb$ anyon around the closed loop generates a phase factor that depends on the anyons inside the loop, see Eq.(\ref{braiding})}
\end{figure}

As shown in Fig.\ref{fig2}, braiding $\eb$ anyon around the closed rectangular area $[4j, 4j+4]\times[x,y]$ gives nontrivial phase factor, reflecting its mutual statistics with other anyons. We first express the local operator for propagation in terms of new variables $\Phi$ and $\Theta$, and count the number of anyons (kinks of $\Phi$) inside the loop by reading how the result depend on $\Phi$ variables. Details are given in the Appendix. We obtain the following result for the braiding phase factor
\beq\label{braiding}
e^{-\frac{i}{2}\pi N(\ob')} \times e^{-i\pi N(\eb')} \times e^{\frac{i}{2}\pi N(\ob)}
\eeq
where $N(\ob)$ stands for the number of $\pi$ kinks of $\Phi_{4j+3}$, and similarly for others. From this result we conclude that the two fundamental anyons $\eb$ and $\ob$ have mutual statistical angle $\pi/2$. Also, as expected from (\ref{enumeration}), $\ob$ and $\ob'$ have different mutual statistics with $\eb$. This again confirms our previous conclusion $\eb\nsim\eb',~ \ob\nsim\ob'$. The self- and mutual statistics of these four elementary anyons are summerized in TABLE \ref{tab1}. 

\begin{table}[h]
\centering
\begin{tabular}{c|cccc}
Statistics & $\eb$ & $\ob$ & $\eb'$ & $\ob'$ \\
\toprule
$\eb$ & 0 & $\pi/2$ &  $\pi$ & $-\pi/2$ \\
$\ob$ & $\pi/2$ & 0 & $-\pi/2$ & $\pi$ \\
$\eb'$ & $\pi$ & $-\pi/2$ & 0 & $\pi/2$ \\
$\ob'$ &  $-\pi/2$ & $\pi$ & $\pi/2$ & 0 \\

\end{tabular}
\caption{\label{tab1}Mutual and self statistical angles between anyons.}
\end{table}

From the mutual statistics between $\eb$ and $\ob$, we conclude that the gapped ground state of our model has $\mathbb{Z}_4$ topological order with fermion parity grading. Anyon contents are
\beq
&\{1, \eb, \eb^2, \eb^3\}\times \{1, \ob, \ob^2, \ob^3\}\times \{1, \psib\}
\eeq
with the fusion rule (\ref{fusion}).

\noindent{\it Anyon Permutation and Symmetry Fractionalization:}  Having established the topological order of our model, we proceed to study how the symmetry acts on anyons. Nontrivial action of symmetry already manifested itself in the braiding result (\ref{braiding}). From the definition of reflection (\ref{reflection}) and anyon creation operators (\ref{creation operator}), we find that:
\beq\label{anyon transformation}
\begin{dcases}
M_{odd}: e^\dagger \rightarrow e^{i\pi/4}e'^\dagger,~o^\dagger \rightarrow e^{i\pi/4}o^\dagger \\
M_{even}: o^\dagger \rightarrow e^{i\pi/4}o'^\dagger,~ e^\dagger\rightarrow e^{i\pi/4}e^\dagger
\end{dcases}
\eeq
Therefore, we conclude that reflection with respect to an odd-numbered wire ($M_{odd}$) changes anyon $\eb$ to $\eb'$, while preserving the anyon type of $\ob$. Likewise, reflection with respect to an even-numbered wire ($M_{even}$) changes anyon $\ob$ to $\ob'$, while preserving the anyon type of $\eb$. 
It is important to note that reflection  interchanges two types of anyons that belong to distinct fermion parity sectors. This kind of anyon permutation by global symmetry cannot occur in  bosonic $\mathbb{Z}_4$ topologically ordered state, where the physical boson corresponds to electron's charge or spin.

We further deduce the fractional quantum numbers of anyons, associated with the $U(1)_{c,a}$ charge/spin conservation and reflection symmetry. From the definition of $U(1)_{c,a}$ symmetry (\ref{reflection}),(\ref{axial}) and anyon creation operator (\ref{creation operator}), we find $\ob$ carries electric charge $1/2$ and spin 0, whereas $\eb$ is charge neutral but carries half of electron's spin.

It follows from the fusion rule (\ref{enumeration}) and the anyon permutation (\ref{anyon transformation}) that the bosonic anyons $\eb^2 = \eb'^2 = \eb^{-2}$ and $\ob^2 = \ob'^2 = \ob^{-2}$ map onto their anti-particles under either reflection $M_{even}$ or $M_{odd}$. In such case, it is meaningful to assign fractional reflection quantum numbers to $\eb^2$ and $\ob^2$ \cite{qi2015detecting}. Unlike physical degrees of freedom that satisfy $M^2=1$, from the action of reflection on the anyon creation operator, we directly obtain $M^2 = -1$ for $\eb^2$ and $\ob^2$. 

\begin{table}[h]
\centering
\begin{tabular}{lc|cccc}
& & $\eb$ & $\ob$ & $\eb'$ & $\ob'$ \\
\toprule
\multirow{2}{*}{} & $M_{even}$ & $\eb$ & $\ob'$ &  $\eb'$ & $\ob$ \\
& $M_{odd}$ & $\eb'$ & $\ob$ & $\eb$ & $\ob'$ \\
\hline
\multirow{2}{*}{}  
& $U(1)_c$ & 0 & 1/2 & 0 & 1/2 \\
& $U(1)_a$ & 1/2 & 0 & 1/2 & 0 \\

\end{tabular}
\caption{\label{tab2}Action of symmetries $M_{even}, M_{odd}, U(1)_c, U(1)_a$ on anyons. First two rows show the permutation of anyon under symmetry, and the last two rows show fractional symmetry quantum numbers of anyons.}
\end{table}

The full pattern of anyon permutation and symmetry fractionalization we have found here constitutes a symmetry anomaly in the fermionic $\mathbb{Z}_4$ topologically ordered surface states of TCIs with $n=2$. We want to stress that the symmetry action on anyons is obtained explicitly and rigorously by microscopic analysis.

\noindent{\it Condensation Transition:}  The anyon permutation property described above has some important implications for the possible phase transition driven by the condensation of anyons. For example, we cannot simply choose to condense $\eb$ anyons alone in the presence of reflection symmetry $M_{odd}$, because $\eb$ maps to $\eb'$ under $M_{odd}$, which is also a bosonic anyon but has nontrivial mutual statistics with $\eb$. The same holds for $\ob$ anyon in the presence of $M_{even}$. When both $M_{even}$ and $M_{odd}$ symmetries are present, two possible condensable anyons are $\eb^2$ and $\ob^2$, which are bosonic anyons and are invariant under anyon permutation. However, condensing any one of them does not completely remove the topological order. Condensing $\eb^2$ for example, leaves $\eb, \ob^2, \eb\ob^2$ deconfined, resulting in $\mathbb{Z}_2$ spin liquid with $U(1)_c$ symmetry but broken reflection symmetry. We cannot remove the topological order completely until we condense at least two anyons successively, which inevitably breaks all the symmetry of the original Hamiltonian.
\begin{figure}
\includegraphics[scale=0.7]{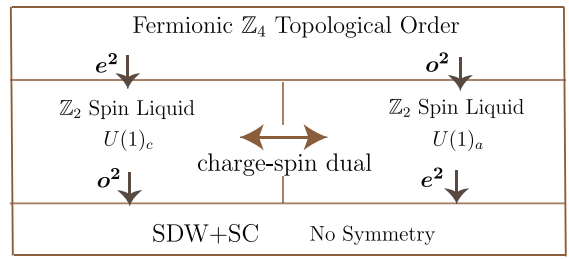}
\caption{\label{fig3} Two possible condensation transitions of anyons in the presence of {\it both} $M_{even}$ and $M_{odd}$.}.
\end{figure}

\noindent{\it 4 and 1 Dirac Cones:}
Lastly, we will remark briefly on the surface topological order in $n=4,1$ Dirac cones. In the case of 4 Dirac cone, as discussed in field theory analysis in \cite{qi2015anomalous}, $\mathbb{Z}_2$ topological order with anomalous symmetry fractionalization is possible. $\Eb$ and $\Mb$ both has fractionalized mirror quantum number $M^2=-1$. Through the microscopic model, we can show this is indeed the case. However, there is no anyon permutation under the reflection symmetry.

For $n=1$ case, we have 1 chiral gapless electron mode on each wire. Though it is not easy to construct topological order purely from the gapless electronic modes, we can use prepared topologically ordered two dimensional plate to facilitate symmetric gapping. By using Bosonic Laughlin state as a decorating plate, we constructed semion-Fermion topological order \cite{fidkowski2013non}. 
Anyon contents are $\{1,\sbb\}\times\{1, \psib\}$, where $\sbb$ is a semion and $\bar{\sbb}$ is anti-semion. This semion Fermion topological order shows elementary form of anyon permutation, where reflection maps $\sbb\leftrightarrow\bar{\sbb}$ and vice versa. 
In the appendix, we present the microscopic construction in detail. 

It is interesting to compare our results for TCIs with reflection symmetry previous works on topological superconductors with charge-conjugation-time-reversal ($CT$) symmetry. Since the simultaneous transformations of reflection and $CT$ is an exact symmetry for any Lorenz-invariant theory, a correspondence between reflection and $CT$ symmetry anomalies is expected when Lorenz invariance can be made emergent. Indeed, our results for $n=1$ and $n=4$ TCIs match with their counterparts in topological superconductors  \cite{metlitski2014interaction,potter2016realizing}.

However, our microscopic model for the case $n=2$ deserves special attention, as it reveals a fundamental difference between reflection and $CT$ symmetry. Here  reflections $M_{even}$ and $M_{odd}$ with respect to two parallel panes at wire $2j$ and  $2j+1$ permute anyons differently. Equivalently, the product of $M_{even}$ and $M_{odd}$, which is spatial translation by one period in the $y$ direction, permute anyon types. This feature has no analog in Hamiltonian-based systems with $CT$ symmetry, as time translation is continuous rather than discrete. It will be interesting to explore possible analogs in periodically-driven Floquet systems.

Our analysis can be extended to other classes of TCIs where two-dimensional surface states can be deformed into an array of coupled wires \cite{ezawa2016hourglass,lu2017classification,huang2017building}. We leave such extension to future works.

\noindent{\it Note Added:} Recently we learned of a related work on surface topological order by Meng Cheng.

\noindent{\it Acknowledgment:}
We thank Yang Qi, Sagar Vijay and Max Metlitski for insightful discussions. This work is supported by DOE Office of Basic Energy Sciences, Division of Materials Sciences and Engineering under Award DE-SC0010526. SH is supported by Samsung Scholarship and Henry Kendall Fellowship from MIT. LF is suppoerted partly by David and Lucile Packard Foundation.

\bibliographystyle{aipnum4-1}
\bibliography{preprint}
\appendix
\setcounter{equation}{0}
\setcounter{figure}{0}
\makeatletter
\renewcommand{\theequation}{S\arabic{equation}}
\renewcommand{\thefigure}{S\arabic{figure}}

\section{\label{appA}Braiding Calculation for Fermionic $\mathbb{Z}_4$ topological order}
In the main text, we presented a local Bosonic operator which makes $\eb$ anyon propagates. In terms of Bosonic operator, effect of that operator is given by the followings.
\beq\label{Bosonic operator}
\Big\{
\begin{array}{ll}
\theta_{4j+1}(x) \rightarrow \theta_{4j+1}(x) -\pi \eta(x-x') \\
\phi_{4j+1}(x) \rightarrow \phi_{4j+1}(x)+\pi \eta(x-x')
\end{array} \nonumber \\
\big\{
\begin{array}{l}
\theta_{4j+2}(x)\rightarrow \theta_{4j+2}(x) +2\pi \eta(x-x')
\end{array} \nonumber \\
\Big\{
\begin{array}{ll}
\theta_{4j+3}(x) \rightarrow \theta_{4j+3}(x) -\pi \eta(x-x') \\
\phi_{4j+3}(x) \rightarrow \phi_{4j+3}(x)-\pi \eta(x-x')
\end{array}
\eeq
Effect of these transformations are $\Phi_{4j}(x) \rightarrow \Phi_{4j}(x)-\pi\eta(x-x') ,~ \Phi_{4j+4}(x) \rightarrow \Phi_{4j+4}(x)+\pi\eta(x-x')$. Following this information, we will present detailed calculation of braiding, and conclude mutual statistics between different anyons (Fig \ref{fig2}).
Here, we will present detailed calculation of braiding $\eb_{4j}$ following the closed loop. We will use the path $(4j,x) \rightarrow (4j+4,x) \rightarrow (4j+4,y) \rightarrow (4j,y) \rightarrow (4j,x)$. This explicitly shows the mutual statistics information between the topological excitations in our system, and gives us hint about the anyon permutation. We can rewrite $\phi$ and $\theta$ in terms of new Bosonic variables.
\beq
\phi_{4j+1} &=& \Theta_{4j+1} ,~ \theta_{4j+1} = (\Phi_{4j+1}-\Theta_{4j} - \Theta_{4j+2})/2 \nonumber \\
\phi_{4j+2} &=& (\Phi_{4j+2}+\Theta_{4j+1}+\Theta_{4j+3})/2 \nonumber \\
\phi_{4j+3} &=& \Theta_{4j+3} ,~ \theta_{4j+3} = (\Phi_{4j+3}-\Theta_{4j+2} - \Theta_{4j+4})/2 \nonumber
\eeq
1) $(4j,x) \rightarrow (4j+4,x)$ is given in the main text. Corresponding operator is
\beq\label{first step phase}
e^{i(-\phi_{4j+1}+\theta_{4j+1})} \times e^{2i\phi_{4j+2}} \times e^{-i(\phi_{4j+3}+\theta_{4j+3})} \nonumber
\eeq
which is equivalent to the following operator
\beq
e^{i\Theta_{4j+1}(x)+i(\Phi_{4j+1}(x)-\Theta_{4j}(x) - \Theta_{4j+2}(x))/2} \nonumber\\
\times e^{i(\Phi_{4j+2}(x)+\Theta_{4j+1}(x)+\Theta_{4j+3}(x))}\nonumber\\
\times e^{-i\Theta_{4j+3}(x)-i(\Phi_{4j+3}(x)-\Theta_{4j+2}(x) - \Theta_{4j+4}(x))/2} \nonumber
\eeq
2) $(4j+4,x) \rightarrow (4j+4,y)$ is easy.
\beq
e^{-i\Theta_{4j+4}(y)+i\Theta_{4j+4}(x)} \nonumber
\eeq
3) $(4j+4,y) \rightarrow (4j,y)$ is similar to 1)
\beq
e^{-i\Theta_{4j+1}(y)-i(\Phi_{4j+1}(y)-\Theta_{4j}(y) - \Theta_{4j+2}(y))/2} \nonumber \\
\times e^{-i(\Phi_{4j+2}(y)+\Theta_{4j+1}(y)+\Theta_{4j+3}(y))}\nonumber\\
\times e^{i\Theta_{4j+3}(y)+i(\Phi_{4j+3}(y)-\Theta_{4j+2}(y) - \Theta_{4j+4}(y))/2} \nonumber
\eeq
4) $(4j,y) \rightarrow (4j,x)$
\beq
e^{i\Theta_{4j}(y)-i\Theta_{4j}(x)} \nonumber
\eeq

Before adding all those, notice that we can consider this operator up to $e^{2i\Theta}$, since this term is a local Bosonic excitation. Also, $\Phi$ s are gapped in the bulk and pinned to the $\langle \Phi \rangle$. Therefore, after adding 1) to 4) all together, we get the total phase factor for braiding.
\beq
&&e^{-i(\langle\Phi_{4j+1}(y)\rangle-\langle\Phi_{4j+1}(x)\rangle)/2} \times e^{-i(\langle\Phi_{4j+2}(y)\rangle-\langle\Phi_{4j+2}(x)\rangle)} \nonumber \\
&&\times e^{i(\langle\Phi_{4j+3}(y)\rangle-\langle\Phi_{4j+3}(x)\rangle)/2} \nonumber \\
&& =e^{-\frac{i}{2}\pi N(\ob')} \times e^{-i\pi N(\eb')} \times e^{\frac{i}{2}\pi N(\ob)}
\eeq
where $N(\ob)$ stands for the number of $\pi$ kinks of $\Phi_{4j+3}$, and similarly for others.

This shows that $\eb$ has mutual semion statistics(statistical angle $\pi$) with $\eb'$, and has statistical angle $\pi/2$ and $-\pi/2$ with $\ob, \ob'$, respectively. Similar calculaion shows that $\ob$ is mutual semion with $\ob'$. $\eb$ and $\ob$ are self Bosons.

\section{\label{appB}Fermion Parity Pattern of Wires in the Ground States of Fermionic $\mathbb{Z}_4$ Topological Order}
We've constructed Fermionic $\mathbb{Z}_4$ topological order in the main text, using the interaction Hamiltonian (\ref{Hv}). Notice that the whole Hamiltonian preserves the Fermion parity of {\it each} wire, and therefore any ground state of the system should have definite Fermion parity for each wire. In the main text, we explicitly constructed a local Bosonic operator which propagates an $\eb$ anyon. Since this operator changes the Fermion parity of wires(though it's a local Bosonic operator in total), topologically protected degenerate ground states are expected to have different Fermion parity patterns.

Suppose our system is placed on the torus(genus$=1$). First check that moving $\eb$ anyon around contractible loop does not change the Fermion parity of any wires. In Fig.\ref{fig2} for example, moving $\eb$ around the loop changes the Fermion parity of $4j+1, 4j+3$ wire twice, therefore preserves Fermion parity of every wire as it should be.

However, following the noncontractible loop makes difference. Suppose for convenience that the total number of wires is a multiple of 4. Group the wires by even indexed, and odd indexed wires. Since we cannot measure the Fermion parity of wires locally, choose one reference ground state and fix every wire's parity as even(so we are caring about the {\it difference} from that reference ground state). Symbol $(+,+)$ means (even/odd) indexed wires both have the same Fermion parity with the reference state.

Define the following noncontrctible loop operators(logical operators which map one ground state to the others) $U_\parallel(\eb)$, $U_\parallel(\ob)$, $U_\perp(\eb)$, $U_\perp(\ob)$. $U_\parallel/U_\perp$ operator first creates quasiparticle-quasihole pair, moves the quasiparticle around the noncontractible loop of torus parallel/perpendicular to the direction of wires, and then pair annihilate again. From the form of the propagation operator, we can see that
\beq
\begin{dcases}
U_\perp(\eb): (+,+)\rightarrow(+,-)\\
U_\perp(\ob): (+,+)\rightarrow (-,+)
\end{dcases}
\eeq
while $U_\parallel$ does not change the Fermion parity. Logical operators satisfy the following algebra,
\beq\label{operator algebra}
U_\parallel(\eb)U_\perp(\ob)=e^{2\pi/4}U_\perp(\ob)U_\parallel(\eb) \nonumber\\
U_\parallel(\ob)U_\perp(\eb)=e^{2\pi/4}U_\perp(\eb)U_\parallel(\ob)
\eeq
while all the other commutators vanish. Each relation in (\ref{operator algebra}) spans 4 dimensional Hilbert space, which in total gives 16 degenerate ground states. Hilbert space spanned by $U_\parallel(\eb), U_\perp(\ob)$ contains 2 basis states with $(+,+)$ pattern, and 2 $(-,+)$ patterns. Similarly, Hilbert space spanned by $U_\parallel(\ob),U_\perp(\eb)$ contains 2 $(+,+)$ patterns and 2 $(+,-)$ patterns. Therefore, we can conclude that among the 16 topologically protected degenerate ground states, we have all possible patterns $(+,+). (+,-), (-,+), (-,-)$, each of them contains 4 states.

\section{\label{appC}$n_M=4$ Surface with $M\times U(1)$ symmetry - Anomalous $\mathbb{Z}_2$ Topological Order}
As an application of the three domain wall interactions we used for $n_M=2$ surface, we will present one way to get $n_M=4$ surface topological order which is anomalous $\mathbb{Z}_2$ topological order. Here we have two pairs(index $a,b$) of helical edge modes in each domain wall, which is described by non-chiral Luttinger liquid with field variables $\phi^{a/b}$ and $\theta^{a/b}$. Define charge($c$) and spin($s$) variables,
\beq
\phi^c \equiv \phi^a+\phi^b,~ \theta^c \equiv \theta^a+\theta^b \nonumber \\
\phi^s \equiv \phi^a-\phi^b,~ \theta^s \equiv \theta^a-\theta^b
\eeq
Commutation relation is
\beq \label{charge commutation relation}
[\partial_x \phi^{c/s}(x), \theta^{c/s}(y)]=2i\pi \delta(x-y)
\eeq
and charge and spin fields commute with each other. We can see the analogy with spinful Luttinger liquid, when we consider $a/b$ as spin degrees of freedom. Free Hamiltonian can be written separated form, by charge and spin sector.
\beq
H_0= H^c_0 (\phi^c,\theta^c, v_c, K_c)+H^s_0 (\phi^s,\theta^s, v_s, K_s)
\eeq
Note that under reflection,
\beq\label{charge sector mirror}
M_j: \Big\{
\begin{array}{ll}
\phi^c_{j+n} \rightarrow \phi^c_{j-n}+ \pi \\
\theta^c_{j+n} \rightarrow \theta^c_{j-n} +(-1)^{j+n}\pi
\end{array}
\eeq
but spin field does not change under reflection.
We will first gap spin sector by {\it intra}-domain wall interaction, in which 4 electrons are involved
\beq\label{spin gap}
H_{spin}=\int dx \sum_j V_s\cos (2 \theta_j^s(x))
\eeq
We can check this is trivially reflection symmetric and also $U(1)$ invariant.

After we gap the spin sector, note that adding or removing a single electron is no longer a low-energy degrees of freedom. Allowed low-energy operators are now adding(removing) Cooper pairs of electrons, or backscattering the Cooper pair of electrons, represented by $e^{2i\phi^c(x)}$ and $e^{2i\theta^c(x)}$  respectively. Remaining gapless charge modes ($H_0^c$) therefore can be considered as a Bosonic quantum wires made of Cooper pairs. Notice that creating a Cooper pair($\sim \exp(i(\phi^c \pm \theta^c))$) creates $2\pi$ kink of $\phi^c$ and $\theta^c$, twice as large as analogous electron case.

We can use same gapping term as we did in $n_M=2$ case in the main text. However, resulting deconfined anyons are a little different, because of the different physical operators allowed. Let's construct three domain wall interactions which gap spin sector as followings. Using redefined variables
\beq\label{charge field redefinition}
&\Phi^c_{2j+1} = \theta^c_{2j}+ 2\theta^c_{2j+1}+ \theta^c_{2j+2} ,~ &\Theta^c_{2j+1} = \phi^c_{2j+1} \nonumber \\
&\Phi^c_{2j} = -\phi^c_{2j-1}+  2\phi^c_{2j}- \phi^c_{2j+1} ,~ &\Theta^c_{2j} =  \theta^c_{2j}
\eeq
the interaction terms we introduce are
\beq
H_V = \int dx \sum_j V_1 \cos 2\Phi_{2j+1} + V_2 \cos 2\Phi_{2j}
\eeq
From (\ref{charge sector mirror}), we can see this interaction term preserves $M$ and $U(1)$, and can also check it can be expressed in physical operators(Cooper pairs).
We will define deconfined anyons $\Mb$ and $\Eb$ as $2\pi$ kink of $\Phi^c_{2j}$ and $\Phi^c_{2j+1}$, created by operators $e^{i \Theta^c_{2j}(x)/2}=e^{i\theta^c_{2j}(x)/2}$ and $e^{i \Theta^c_{2j+1}(x)/2}=e^{i\phi^c_{2j+1}(x)/2}$, respectively.
Note that $\pi$ kinks are {\it not} a deconfined anyons here, since there's no physical operator without energy cost that makes $\pi$ kink propagate. In comparison, $2\pi$ kink can propagate freely, by the following transformation
\beq\label{gauge transformation combination}
\Big\{
\begin{array}{ll}
\theta^c_{j}(x) \rightarrow \theta^c_{j}(x) +2\pi \eta(x-x') \\
\phi^c_{j}(x) \rightarrow \phi^c_{j}(x)+2\pi \eta(x-x')
\end{array}
\eeq
which moves $2\pi$ kink at $j-1$ to $j+1$. Note that $4\pi$ kinks can be made from local Bosonic operators $e^{2i\theta^c_{2j}(x)}$ and $e^{2i\phi^c_{2j+1}(x)}$, which is therefore trivial.

Therefore, $\Eb^2 \sim \Mb^2 \sim 1$. Mirror quantum numbers are
\beq\label{correct mirror}
M^2_{\Eb}=-1,~M^2_{\Mb}=-1
\eeq
which is anomalous pattern and only realizable on the surface of 3D. $\Eb$ has $U(1)$ charge 1, while $\Mb$ has zero. So in conclusion, we get anomalous $\mathbb{Z}_2$ topological order. This agrees the result of \cite{qi2015anomalous}, where authors use field theoretical approach (double-vortex condensation) to get the same result. Note that we do not have anyon permutation here.

\section{\label{appD}$n_M=1$ Surface with $M$ symmetry - Semion Fermion Topological Order}

\begin{figure}
\includegraphics[scale=0.8]{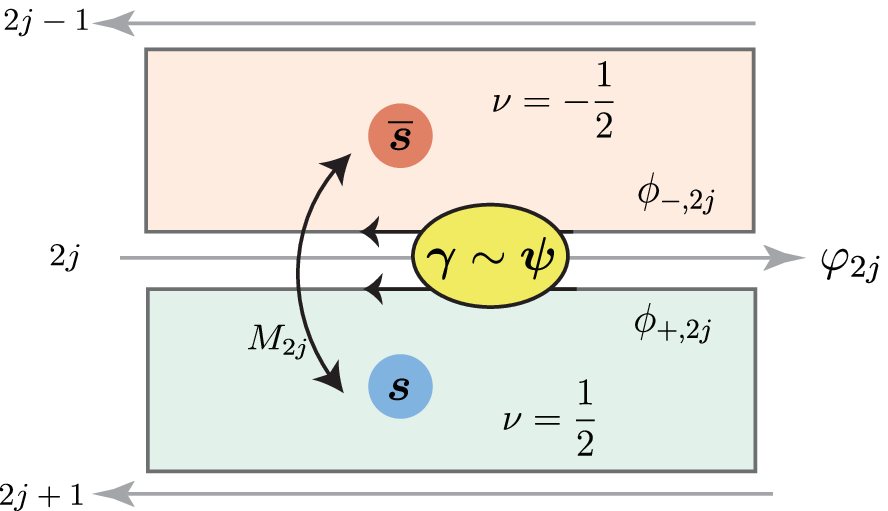}
\caption{\label{figS1} Schematic description of constructing reflection preserving semion-Fermion topological order on the $n_M=1$ surface. $\sbb$ and $\bar{\sbb}$ are semions and anti-semions respectively. $\psib$ is a physical electron, which can be identified to emergent neutral Majorana Fermion.}
\end{figure}
When $n_M=1$, we have single surface Dirac cone, and therefore each domain wall carries single {\it chiral} electron mode. Gapping term constructed purely from given chiral electron modes is hard to construct. Better strategy for symmetric gapping is to add additional degrees of freedom by depositing 2D topologically ordered systems between domain walls, while preserving all symmetries of the system. This strategy is indebted to the work \cite{mross2015composite}. We will use Bosonic quantum Hall plates with $K$ matrix $K=\pm (2)$, with fundamental Boson $\rho$ (This Boson can be considered as paired electrons). Note that the decorating plate itself can be easily constructed microscopically \cite{teo2014luttinger}.  As a result, we will get semion-Fermion topological order with reflection symmetry, first suggested in \cite{fidkowski2013non} for in time-reversal case. Here, importantly, we don't have symmetry fractionalization of $M$. However, we still have elementary anyon permutation, which might potentially captures the symmetry anomaly on the surface \cite{fidkowski2015realizing}. This again stress the importance of understanding anyon permutation.

To preserve the $M$, we will deposit 2D system with $\nu=1/2$ and $\nu=-1/2$ alternatively, as in the Fig.\ref{figS1}. This process adds 2 copropagating gapless bosonic modes($\phi_{+,j}, \phi_{-,j}$, with $\rho \sim e^{2i\phi_{\pm}}$) at each interface, propagate oppositely to the chiral electron mode(let's say $\varphi_j$, with $\psi \sim e^{i\varphi}$).

We have the following commutation relations of gapless modes at interface $j$:
\beq
&[\partial \varphi_j(x), \varphi_j(x')]=(-1)^j 2\pi i \delta(x-x') \nonumber \\
&[\partial \phi_{\pm,j} (x), \phi_{\pm,j}(x')]=(-1)^{j+1}\pi i \delta(x-x')
\eeq
with reflection $M$ act on them as
\beq\label{mirror reflection actions}
M_j:
\begin{dcases}
\varphi_{j+n} \rightarrow \varphi_{j-n} +\frac{\pi}{2} (1+(-1)^{j+n+1})\\
\phi_{+,j+n} \leftrightarrow \phi_{-,j-n}
\end{dcases}
\eeq
Therefore, $e^{i \phi_{\pm}}$ creates an edge semion with statistical angle $\pi/2$. Note that semion modes propagate opposite direction from the chiral electron mode.

Next step is to introduce the following physical interaction, while preserving $M$.
\beq
H_c&=& \int dx \sum_j v_c [(\psi_j)^2 (\rho_{+,j}^\dagger)(\rho_{-,j}^\dagger)+h.c.] \nonumber\\
&=&\int dx \sum_j v_c \cos 2\Theta_j
\eeq
with redefined Bosonic field $\Theta$ and conjugate $\Phi$ defined as
\beq
&\Theta_j \equiv \varphi_j - \phi_{+, j} - \phi_{-, j},~\Phi_j \equiv \varphi_j + \phi_{+, j} + \phi_{-, j} \nonumber \\
&[\partial_x \Theta_j(x), \Phi_j(x')]=4\pi i \delta(x-x')
\eeq

Therefore, microscopically, this interaction annihilates two $\varphi_j$ electrons from the chiral electron mode and adds one boson each to the left/right side decorating plate. From (\ref{mirror reflection actions}), this interaction preserves $M$. Note that $4\pi$ kink of $\Theta_j$ corresponds to trivial anyon, because applying local Bosonic operator $(\psi_j)^2\sim e^{2i\varphi_j}$ makes it.

There's still a gapless degree of freedom left in each domain wall, which comes from nonzero total central charge(remaining thermal Hall conductivity) of each domain wall. This remaining gapless mode is given by
\beq
\phi^n_j =\phi_{+,j}-\phi_{-,j}
\eeq
From the commutation relation of $\phi_{\pm,j}$, we can see $\psi \sim e^{i \phi^n_j}$ is a Fermion. This emergent Fermion is constructed entirely from the semionic edge modes of Boson quantum Hall states, with zero overlap with the physical fermion. However, importantly, this emergent Fermion is {\it not} a fractional excitation, since we can construct following operator,
\beq\label{physical}
e^{i\phi^n_j}\times e^{\pm i\Theta_j} &=& e^{ \pm i\varphi_j} e^{\mp 2i\phi_{\mp,j}}
\eeq
which is physical. Since $\Theta$ is gapped by $H_c$, we conclude that the emergent Fermion is topologically equivalent to the physical electron ($e^{i(\phi_{+,j}-\phi_{-,j})} \sim e^{i\varphi_j}$), which we will say $\psib$.

We now split the emergent Fermion operator $\psi^n$ into two Majorana operators.
\beq\label{majorana definition}
\psi^n_j = e^{-i \pi/4} (\gamma_{j,L} + i \gamma_{j, R})
\eeq
Majorana operators satisfy $\gamma^2=1$.  We will add 2-Majorana interaction term as follows
\beq\label{semion fermion majorana}
H_{int}= iV\sum_{j} \int dx \gamma_{j, R} \gamma_{j+1, L} \cos (\Theta_j - \Theta_{j+1})
\eeq
where $V$ is real constant. Under $M_j$, $\phi^n_j \rightarrow -\phi^n_j$ which leads to $\gamma_L \leftrightarrow \gamma_R$ from (\ref{majorana definition}). Therefore, $H_{int}$ is reflection symmetric.
Importantly, it follows from (\ref{physical}) that $H_{int}$ consists of physical electron and boson operators only. $\Theta$ depending factor must be included, in order to make the interaction physical. Full Hamiltonian is given  by $H_0+H_c+H_{int}$, and this completely gaps the $n_M=1$ surface(except the edge).

Next, let's analyze the fractional excitations and the resulting topological order. Apparently, there are 4 possible candidates for such excitations, kinks of $\Theta$, (anti-)semion excitations from decorating plate, and emergent Fermion $\psi^n$(or equivalently $\gamma$). Let's start from the smallest kink excitation of $\Theta_j$, which is $\pi$ kink. There's no physical operator which creates any $\pi$ kinks of $\Theta$ without affecting emergent Fermion sector. However, we claim that the combination ($\pi$ kink of $\Theta$)+($\pi$ kink of $\phi^n$) is deconfined excitation, which is created by simply $e^{i\phi_{+,j}}$. This is just a semion $\sbb$ ($j$ is even) or anti-semion $\bar{\sbb}$($j$ is odd) in the decorating plate. So it's enough to see whether (anti-) semion can cross the interfaces. Now, the following operator removes an (anti-)semion from the plate between $(j,j+1)$ and creates an (anti-)semion in the plate $(j+2, j+3)$ plate.
\beq\label{semion propagation}
(e^{-i\phi_{-,j+1}}e^{i\phi_{+,j+1}})(e^{-i\phi_{-,j+2}}e^{i\phi_{+,j+2}})
\eeq
This operator is a local Bosonic operator, since each parenthesis is equivalent to physical electron operator and we have {\it even} number of it. We conclude that $\sbb$ and $\bar{\sbb}$ are deconfined anyons excitations. They differ from each other by emergent Fermion($\sim$ physical electron) $\psib$, so we have a fusion rule $\bar{\sbb}\sim \sbb \times \psib$

We should check whether this excitation is accompanied by additional structure coming from the emergent Fermion sector, since Majorana interaction terms (\ref{semion fermion majorana}) contains $\Theta$ dependent factor. However, we can see that the change of $\cos (\Theta_j -\Theta_{j+1})$ exactly compensates with the change of $\gamma$, and therefore the (anti-)semionic excitation does not change the form of (\ref{semion fermion majorana}). In other words, it does not bind any non-Abelian structures(zero mode).

Therefore, full anyon contents can be expressed as $\{I,\sbb\} \times \{I, \psib\}$, with fusion rules are the followings
\beq
\begin{dcases}
\sbb \times \sbb \sim I \nonumber \\
\psib \times \psib \sim I
\end{dcases}
\eeq
This is semion-Fermion topological order, which was originally proposed in time-reversal invariant case by stacking 2D layers and condense anyon sets \cite{fidkowski2013non}.

The action of $M$ shows anyon permutation
\beq
M:
\begin{dcases}
\sbb \leftrightarrow \sbb\psib (=\bar{\sbb}, \text{anti-semion}) \\
\psib \leftrightarrow \psib~(\text{invariant})
\end{dcases}
\eeq
but there's no symmetry fractionalization since $M^2$ acts trivially both to physical electrons and physical Bosons. So symmetry anomaly might be potentially detected only from anyon permutation pattern.

Edge modes can be read from the Hamiltonian easily. There are 2 counter-propagating gapless modes on the leftmost side of the surface, which commute with all terms in the Hamiltonian. One is gapless Majorana, $\gamma_{1,L}$, and the other is semionic mode from decorated plate. Notice that although we have Majorana edge mode, our topological order is Abelian. This is however natural, because our emergent Majorana Fermion is topologically equivalent to the physical electron.

\end{document}